**Hexagonal boron nitride/bilayer graphene moiré superlattices in the Dirac-material family: energy-band engineering and carrier doping by dual gating**

Takuya Iwasaki[1,*], Yoshifumi Morita[2,*]

[1] *Research Center for Materials Nanoarchitectonics, National Institute for Materials Science (NIMS), 1-1 Namiki, Tsukuba, Ibaraki 305-0044, Japan*
[2] *Faculty of Engineering, Gunma University, Kiryu, Gunma 376-8515, Japan*

*E-mail: IWASAKI.Takuya@nims.go.jp (T.I.), morita@gunma-u.ac.jp (Y.M.)

**Abstract**

We review the fabrication and transport characterization of hexagonal boron nitride (hBN)/Bernal bilayer graphene (BLG) moiré superlattices. Due to the moiré effect, the hBN/BLG moiré superlattices exhibit an energy gap at the charge neutrality point (CNP) even in the absence of a perpendicular electric field. In BLG, the application of a perpendicular electric field tunes the energy gap at the CNP, which contrasts with single-layer graphene and is similar to the family of rhombohedral multilayer graphene. The hBN/BLG moiré superlattice is associated with non-trivial energy-band topology and a narrow energy band featuring a van Hove singularity. By employing a dual-gated device structure where both the perpendicular displacement field and the carrier density are individually controllable, systematic engineering of the energy-band structure can be achieved. The data presented here demonstrate the universality and diversity in the physics of hBN/BLG moiré superlattices.

# 1. Introduction

Carbon-based superlattices are promising from the perspective of low-dimensional quantum metamaterials. Possible candidates for the templates of such superlattices are low-dimensional carbon materials such as carbon nanotubes [1] and graphene [2–4]. When atomic-layer materials with similar lattice structures are stacked, long-range spatial modulation can occur, known as a moiré pattern. A stack exhibiting a moiré pattern, i.e., a moiré superlattice, displays new emergent properties that do not exist in the parent materials. These emergent properties arise from the modulated/reconstructed band structure, which depends on the combination and stacking (twist) angle of the original materials [5–12]. In particular, the modulation of the band structure of graphene using hexagonal boron nitride (hBN, Fig. 1(b)) offers an unprecedented route to novel energy-band engineering [11–16]. Single-layer graphene (SLG) exhibits a relativistic nodal energy dispersion of $E \sim k$ [3,4]. On the other hand, Bernal (AB) stacked bilayer graphene (BLG, Fig. 1(a)) and rhombohedral multi($N$-)layer graphene show their degeneracy as $E \sim k^N$ (and its modified form) in the energy dispersion [17,18]. Interaction-driven symmetry breaking is expected due to the touching of such degenerate energy states with a finite (or divergent) density of states. Importantly, the application of a perpendicular electric field tunes the energy gap, in contrast to SLG [19,20]. This systematic engineering of the energy-band structure and carrier doping was demonstrated in hBN/BLG moiré superlattices with an alignment angle of nearly zero degrees [21]. In recent studies, a "dual-gated" device structure has played a key role in the discovery of a variety of intriguing quantum phenomena in hBN/BLG moiré systems [21–27], as well as in BLG systems lacking alignment to hBN [28–37]. These parent materials (and devices) are foundational to the "Dirac-material family," which is linked to rhombohedral multilayer graphene [38–41].

The early days of hBN/graphene moiré superlattices began with studies using scanning tunneling microscopy/spectroscopy [13]. For a review of the early developments in hBN/graphene heterostructures, please refer to [11] and its updated version [12]. This review focuses on the transport properties of hBN/BLG moiré devices. This review is organized as follows: Section 2 provides a brief overview of the latest research on BLG and hBN/BLG moiré systems; Section 3 describes the device fabrication techniques; Section 4 discusses the transport properties of dual-gated hBN/BLG moiré devices; and Section 5 offers a summary and future prospects.

# 2. Electronic structures of BLG and hBN/BLG moiré systems
## 2-1. High-quality graphene devices

Graphene, a monolayer of carbon atoms, was discovered early in this century when it was exfoliated on a SiO$_2$/Si substrate [2]. This famous "Scotch-tape" method is now well known beyond the graphene community. However, since every atom in graphene resides on the SiO$_2$ surface, the basic properties of graphene are highly sensitive to the environment. The SiO$_2$ surface is not atomically flat and contains dangling bonds, which are responsible for surface charge traps and charge-impurity doping (see the left image in Fig. 1(c)). Furthermore, the surface optical phonon of $SiO_2$ (~59 meV [42,43]) has lower energy compared to hBN (>160 meV [44]), which causes difficulties in the device performance [45]. Progress was made by suspending graphene, which improves device quality (see ref. [46] for a benchmark). However, the geometry of suspended graphene poses severe limitations on the device architecture.

Currently, we are in the age of "after hBN," where hBN-encapsulated structures, i.e., hBN/graphene/hBN, overcome such difficulties. hBN is an atomically flat two-dimensional (2D) material in which boron and nitrogen atoms are arranged in a honeycomb lattice similar to that of graphene (Fig. 1(b)). hBN has a large band gap (~6 eV) and is expected to be free of dangling bonds and surface charge traps [47,48]. Combined with the edge-contact technique [49], the graphene is effectively decoupled from the metallization and is only coupled to hBN during the fabrication process, which helps keep the device away from scattering sources typically encountered in conventional processes. Additionally, by employing a graphite gate structure, ultra-clean graphene devices are now available. The typical fabrication process for such devices will be described in Section 3.

### 2-2. Dual-gated BLG systems

Following the discovery of SLG, BLG was first identified in 2006 [50]. BLG consists of two SLG sheets stacked in a thermodynamically stable structure known as Bernal or AB stacking, where the carbon atoms of the *A*-sublattice in the upper layer overlap with the carbon atoms of the *B*-sublattice in the lower graphene layer (Fig. 1(a)). As illustrated in Fig. 2(a), the electronic band structure of BLG has parabolic conduction and valence bands that touch at the $K$ ($K'$) point in the first Brillouin zone (BZ) [51,52]. Electrons in the graphene family possess spin and valley degrees of freedom, often referred to as "flavor." The contact point between the conduction and valence bands is called the charge neutrality point (CNP). Breaking the spatial inversion symmetry of BLG opens a gap at the CNP. This can be achieved by applying an electric field perpendicular to the BLG plane [19,20] and/or by stacking BLG with hBN at an angle close to ~0° to form a moiré superlattice [14,21,37,53–56]. Opening a gap at the CNP induces a finite Berry curvature [57], which leads to the generation of topological flavor currents that are transverse to an in-plane electric field [35–37,57–59] (as detailed in Section 4-4). At the band edge, the Fermi surface splits into three side pockets and one inverted center pocket (Fig. 2(b)) due to the trigonal warping effect (Fig. 2(c)), resulting in a divergent density of states (i.e., van Hove singularity: vHs) and an abrupt change in the topology of the Fermi surface (i.e., Lifshitz transition) [33,51,52].

In recent years, the study of BLG under a perpendicular electric field has revealed its rich phase diagrams and quantum phenomena. These include the fractional quantum Hall effect with even denominators [28–30], flavor symmetry breaking away from the CNP [31–34], orbital ferromagnetism and anomalous Hall effect [32], half and quarter metal phases (Fig. 2(d)) [33], and superconductivity induced by in-plane magnetic fields (Figs. 2(e-g)) [34]. Notably, Zhou *et al*. [34] demonstrated a cascade of phase transitions characterized by different Fermi pocket occupations alongside flavor symmetry breaking (Fig. 2(f)). These phases were discovered by exploring the parameter space, particularly through the independently control of the perpendicular electric field and carrier density. It is important to note that applying an electric field perpendicular to the BLG plane significantly contributes to layer polarization and flavor symmetry breaking in the energy states near the CNP and vHs.

### 2-3. hBN/BLG moiré systems

When the stacking angle between hBN and BLG approaches approximately 0°, a long-period (~14 nm) moiré pattern is formed (as illustrated in the right image in Fig. 1(c), which shows a moiré pattern of hBN/SLG for simplicity. For the relation between the moiré wavelength and the stacking angle, see Fig. 7(g)). To calculate the hBN/BLG band structure at an arbitrary stacking angle, Moon and Koshino [53] proposed an effective continuum model. In this model, the effective Hamiltonian, expressed in the basis corresponding to the wave functions $\Psi = (\psi_{A1}, \psi_{B1}, \psi_{A2}, \psi_{B2})$ (refer to Fig. 1(a) for the arrangement) of the hBN/BLG moiré superlattice near the $K$ point, is written as follows [53]:

$$\mathcal{H} = \begin{pmatrix} H_G + V_{hBN} & U_{BLG}^\dagger \\ U_{BLG} & H_G \end{pmatrix} \quad (1)$$

where $H_G \approx -\hbar v \mathbf{k} \cdot \boldsymbol{\sigma}_\xi$ is the Hamiltonian of SLG near the $K_\xi$ point, $\xi = \pm 1$ the valley index, $\hbar = h/2\pi$, $h$ Planck's constant, $v$ the band velocity, $\mathbf{k}$ the wavevector, and $\boldsymbol{\sigma}_\xi = (\xi\sigma_x, \sigma_y)$ Pauli matrices. $U_{BLG}$ is the interlayer coupling between two SLG sheets in the AB stack [51,52]:

$$U_{BLG} = \begin{pmatrix} 0 & \gamma_1 \\ -\hbar v_3 (\xi k_x - k_y) & 0 \end{pmatrix} \quad (2)$$

where $\gamma_1$ (0.30–0.404 eV [52,53,60–62]) is the coupling of B1-A2 vertically stacked dimer, and $v_3$ ($v_3 = \sqrt{3} a_{gr} \gamma_3 / 2\hbar$ where $a_{gr} \sim 0.246$ nm is the lattice constant of graphene, $\gamma_3 = 0.10$–0.38 eV [52,53,60–62] is the coupling of A1-B2 site, corresponding to $v_3 = 0.03$–$0.12 \times 10^6$ m/s) is the effective velocity due to the A1-B2 weak coupling, which is responsible for the trigonal warping effect. $V_{hBN}$ represents the effective potential due to hBN [53] (such that the BLG band structure is described by Eq. 1 without $V_{hBN}$). The band structure of the hBN/BLG moiré superlattice at a stacking angle of $\theta = 0°$, calculated by this model, is shown in Fig. 3(a) [53]. An energy gap of ~40 meV opens at zero energy (i.e., the CNP) due to the absence of the inversion symmetry in $V_{hBN}$. Consequently, the hBN/BLG moiré system lacks inversion symmetry. The gap also opens at the corners of the mini BZ (Fig. 3(b)), referred to as the satellite points, between the main band and both the electron-side and hole-side remote bands. The gap on the electron side is smaller than that on the hole side. This electron-hole asymmetry arises from the difference in the phases of the matrix elements in the model, which result from the mixing of the original energy dispersion and the effective potential of $V_{hBN}$ [53]. As the gap opens at the CNP and satellite points, the bandwidth of the main band narrows, leading to the emergence of vHs between the CNP and the satellite points. Pantaleón *et al*. [54] calculated the hBN/BLG moiré band structure, demonstrating that the bandwidth and the Chern number of the main band can be tuned by a perpendicular displacement field (Figs. 3(c,d)).

In hBN/BLG moiré devices, the Hofstadter butterfly spectrum has been observed due to the interaction between the electrical potentials (moiré unit cell) and magnetic fields (cyclotron orbits) at comparable length scales [14,53,63]. Recently, intriguing quantum phenomena have been reported, including the fractional Chern insulator under large magnetic fields [22], the valley Hall effect [58,59], unconventional ferroelectricity [23,25] and the ratchet effect [24,26], and Umklapp electron-electron scattering [64–66]. In particular, transport studies using THz excitation (Figs. 3(e-g)) [66] and de Haas-van Alphen spectroscopy via the scanning superconducting quantum interference device-on-tip technique (Figs. 3(h-k))[67] have revealed details of the hBN/BLG moiré band structure expanded to a high energy range, where multiple Fermi pockets coexist (Fig. 3(j)), leading to a compensated semimetal phase [66] and coherent magnetic breakdown (Fig. 3(k)) [67]. These unique phenomena and tunable band structures suggest that the hBN/BLG moiré system provides an ideal platform for studying correlated and topological properties, similar to those observed in twisted bilayer graphene (TBG) at the "magic-angle" [5–9] and rhombohedral multilayer graphene family [38–41]. In contrast to TBG systems, the hBN/BLG moiré superlattice at a stacking angle close to 0° is thermodynamically stable due to its commensurate state [59]. However, the precise angle alignment between hBN and BLG flakes is not as well established as in the TBG stacks, where the same graphene flake can be assembled using the "tear-and-stack" [68] or "cut-and-stack" method by atomic force microscopy (AFM) [69] or laser ablation [70]. We will discuss ways to overcome this issue in Section 3-2.

## 3. Device fabrication process

Independent control of an electric (displacement) field and carrier density can be achieved in a dual-gated device with top and back gates. With the exception of a few devices [32], this has been accomplished using high-quality dual-gate BLG devices based on hBN/BLG/hBN heterostructures. In this Section, we describe our latest device fabrication process and some state-of-the-art techniques.

### 3-1. Dry transfer technique

First, using the Scotch tape method, BLG and hBN flakes are exfoliated from highly-oriented-pyrolytic graphite or Kish graphite and hBN bulk crystals to Si substrates with $SiO_2$ layers of ~90 nm or ~300 nm thickness to enhance the optical contrast of the flakes [71,72]. BLG flakes can be identified by their optical contrast [73] and Raman spectroscopy [74,75]. AFM can be used to measure the thickness of hBN. On $SiO_2$/Si substrates, atomically flat surfaces are consistently achieved when each hBN flake exceeds ~15 nm in thickness [47]. In fact, hBN/graphene/hBN heterostructures with the bottom hBN layers thicker than >15 nm exhibited markedly higher carrier mobility [76]. Moreover, hBN effectively screens the charge-potential inhomogeneities induced by charged impurities in the underlying SiO$_2$ substrate [77,78], and this screening capability increases with the number of hBN layers [79,80].

To assemble a stack, the dry transfer technique using a polymer stamp is widely employed [49,81,82]. Commonly used polymers for the stamp include polydimethylsiloxane (PDMS), polycarbonate (PC), and polypropylene carbonate (PPC). The most popular combination is PDMS/PC, where PDMS serves as a cushion and PC acts as an adhesive for the 2D materials. In the typical procedure, the PDMS/PC stamp is brought into contact with a 2D material, picked up, and then brought into contact with another 2D material and picked up again. Once the desired stack is assembled by repeating this process, the stack with the PC film is dropped onto a SiO$_2$/heavily doped Si substrate (dropping only the stack is challenging for the PDMS/PC stamp due to the strong adhesion of PC). Afterward, the PC residue is removed by dipping in chloroform and then optionally annealing in an Ar/H$_2$ gas atmosphere. For further cleaning, mechanical squeezing using the AFM contact mode can be employed to eliminate interfacial bubbles and surface contamination [83].

In our transfer method, as shown in Fig. 4(a) [84], the PDMS/PPC stamp is used to pick up the hBN flake, which is dropped onto the BLG (unlike the PDMS/PC stamp, the PDMS/PPC stamp can only drop a 2D material/stack). The drop process is performed at ~110°C, with a slow contact speed and a finite angle between the picked-up stack and the substrate, which helps suppress bubble formation at the 2D interface. This pick-up/drop process is repeated to create the hBN/BLG/hBN stack.

To use a graphite flake as the back-gate electrode, the hBN/BLG/hBN is picked up and dropped onto a graphite flake with a "desired shape" which should extend partially outside of the BLG flake (as detailed in Section 3-3). The graphite back gate can reduce the influence of charged impurities and potential inhomogeneity in SiO$_2$ [12,16]. Furthermore, the graphite back gate has been found to be more effective than other metallic options, which are associated with greater interface contamination and possible grain boundaries [28,85]. If graphite is used for both top- and back-gate electrodes, the best device quality could be achieved; however, one should be more cautious about interfacial bubbles and the shape of each flake. The completed stack is annealed in an Ar/H$_2$ gas atmosphere to remove any possible PPC residue. An example of the hBN/BLG/hBN/graphite stack is shown in Figs. 4(b-f).

### 3-2. Angle alignment of moiré superlattices

The stacking angle between 2D materials is controlled during the transfer process by rotating the substrate. In the case of making a TBG stack, the same single-crystal SLG flake can be used through the tear(cut)-and-stack method [68–70], which ensures an accurate relative twist angle. This is not the case for stacks composed of different crystals, as the crystal orientation of the as-exfoliated flakes is random. For the fabrication of hBN/BLG moiré superlattices, it is essential to determine the crystal orientation of both hBN and BLG flakes. Typically, a long straight edge of a flake is considered to represent the crystal plane, as it tends to appear there (Fig. 5(a)) [4,86]. However, graphene and hBN possess two types of crystal planes known as armchair and zigzag structures, respectively, which are misaligned by $30° + 60i°$ ($i$ is an integer; see Fig. 1(b); the edge along the x-axis is zigzag, while the y-axis is armchair). The crystal orientation of the graphene edge can also be identified through spatial mapping Raman spectroscopy, where the D band intensity is pronounced at the armchair edge (Fig. 5(b)) [86]. For hBN (and non-centrosymmetric 2D materials), the laser polarization dependence of second harmonic generation (SHG) can be utilized to determine the crystal orientation (Fig. 5(c)) [87,88]. Recent studies have demonstrated that one of the ideal tools for precisely and quickly identifying crystal orientation is torsional force microscopy (TFM), a type of scanning probe microscopy sensitive to dynamic friction [89]. TFM can visualize the atomic lattice of graphene and hBN on $SiO_2$ or a polymer stamp in an ambient atmosphere (Figs. 5(d,e,g)). After stacking, annealing the stack at temperatures above 200°C can promote the stacking angle between hBN and graphene to approach ~0° or ~30°, depending on the initial stacking angle (Figs. 5(h-j)) [90–92].

Eventually, the stacking angle can be confirmed through transport measurements at low temperatures (as detailed in Section 4-1). However, since completing device fabrication and conducting low-temperature transport measurements can be time-consuming, it is desirable to verify the intended stacking angle soon after assembling the stack. To this end, techniques such as TFM [83,88], lateral/friction force microscopy [93,94], and piezoresponse force microscopy [94,95] can be employed to image the moiré pattern immediately after picking up the stack with a polymer stamp (Figs. 5(f,g)). The stacking angle observed at this stage is not expected to change significantly during the subsequent standard fabrication process [88]. For the completed stack, SHG can also be used to identify the crystal orientation of the moiré pattern and each layer [24,96,97].

### 3-3. Etching and metallization

In the early days, electrical contacts were made by directly depositing metals onto the graphene surface ("surface contact"). However, graphene lacks surface bonding sites, which results in high contact resistance [98]. To improve the metal-graphene contact, we employ the graphene-metal "edge contact" [49], as described below.

Figure 6(a) illustrates the schematic flow and corresponding optical images of the fabrication process for the early devices. The completed stack (hBN/BLG/hBN/graphite) is patterned into a Hall bar geometry using electron beam lithography (EBL) and reactive ion etching (RIE) with fluorinated gases ($CHF_3$, $CF_4$, $SF_6$, etc., for etching hBN) and oxygen plasma (for etching graphene). After patterning, the edge of the BLG is exposed between the hBN layers. Contact electrodes are then deposited on the edge of the BLG and the bottom graphite using second EBL and electron beam evaporation. Typically, Au/Cr or Au/Pd/Cr are used as electrode metals [49]. Following the contact fabrication, an additional hBN flake is transferred onto the stack to cover the edge of the BLG, resulting in a stack configuration of hBN/hBN/BLG/hBN/graphite. This additional hBN layer is incorporated to prevent leakage between the top-gate and the edges of the BLG channel. Finally, the top-gate electrode is deposited using third EBL and evaporation. If graphite is used for both top and back gates, the top-gate fabrication steps (iv and v) are not necessary.

Figure 6(b) depicts the improved method, in which the top-gate electrode is first deposited onto the completed stack using EBL and evaporation. Next, EBL and RIE are performed to pattern the Hall bar geometry. Following this, the contact electrodes are fabricated through EBL and evaporation. Compared to the previous procedure, this method avoids the need to transfer an additional hBN layer, which can lower the top-gate capacitance. An alternative approach for fabricating the contact electrodes involves performing RIE and evaporation using the same mask to prevent contamination between the metal and the BLG. In this case, an additional EBL and RIE step is required to pattern the stack into the Hall bar geometry.

When graphite is used as the gate electrode, the Hall bar contact electrodes are typically designed to be positioned outside the graphite flake to avoid leakage. Figure 4(f) shows an example of the hBN/BLG/hBN/graphite stack, where the graphite flake is partially outside the BLG (and vice versa). When employing graphite for both the top and back gates, this condition should be met for both graphite flakes. In the case of a local gate structure, while the BLG near the contact regions cannot be gated by the local gate, it can be independently tuned using the heavily doped Si back gate. Gating the entire region with contacts nearby using the local back gate without causing a short circuit can be achieved by fabricating the contacts through selective etching of the top hBN (with $CF_4$ [99], $XeF_2$ [100], $SF_6$ plasma [101]) or by using an etch stop immediately after exposing the edge of the BLG [102].

## 4. Transport properties in hBN/BLG moiré devices

In this Section, we discuss the transport properties of dual-gated hBN/BLG moiré superlattices, primarily exemplified by our devices [21,37]. The schematic cross-section and optical images of the main devices (D1 [21], D2 [37]) are shown in Figs. 7(a-c). Their fabrication processes are shown in Figs. 6(a,b), respectively. D1 consists of a top-gate(Au/Ti)/hBN/hBN/BLG/hBN/graphite back-gate stack. D2 is composed of a top-gate(Au/Ti)/hBN/BLG/hBN stack, with the $SiO_2$/Si substrate serving as the back gate (note that the stacking angle between hBN and BLG in D2 is not close to 0°, but also not too far). A four-terminal method with AC lock-in techniques (an excitation current $I \sim 10–100$ nA and a frequency of ~17 Hz) was employed to measure the longitudinal resistivity $\rho_{xx} = (V_{xx}/I)(W/L)$ and the Hall resistivity $\rho_{xy} = V_{xy}/I$, where $L$ and $W$ are the channel length and width, respectively, $V_{xx}$ and $V_{xy}$ are the voltages measured between the Hall-bar electrodes (Fig. 7(b)). The devices were cooled to low temperatures ($T$) in a $^4$He cryostat equipped with a superconducting magnet to apply a magnetic field ($B$) perpendicular to the device plane.

### 4-1. Dual-gating transport

Here, we present the basic transport properties and the estimation of the moiré period and stacking angle in D1. Figure 7(d) displays the dual-gate map of $\rho_{xx}$ for D1 as a function of $V_{tg}$ and $V_{bg}$ at $T = 1.6$ K. Through this dual gating, the total carrier density ($n$) and the average perpendicular displacement field ($D$) can be controlled independently. These are defined as follows:

$$n = \frac{C_{bg}V_{bg} + C_{tg}V_{tg}}{e} + n_r \tag{3}$$

$$D = \frac{C_{bg}V_{bg} - C_{tg}V_{tg}}{2\varepsilon_0} + D_r \tag{4}$$

where $C_{bg(tg)}$ is the back-(top-)gate capacitance per unit area, $e$ the electron charge, and $\varepsilon_0$ the vacuum permittivity. $n_r$ and $D_r$ are the residual carrier density and displacement field, respectively. Note that, as defined above, a single-gated device structure cannot modulate one of these parameters independently. Figure 7(e) presents the $n$-$D$ map converted from Fig. 7(d). Figure 7(f) shows $\rho_{xx}$ ($B = 0$) and $\rho_{xy}$ ($B = 0.2$ T) for D1 as a function of $n$. The CNP at $n \sim 0$ is characterized by a peak in $\rho_{xx}$ and a sign change in $\rho_{xy}$. The peaks in $\rho_{xx}$ around $n \sim \pm 2.4 \times 10^{12}$ cm$^{-2}$ correspond to the satellite points of the moiré band (the corners of the mini BZ). Starting from the CNP, when electrons or holes are doped, the carrier type (the sign of $\rho_{xy}$) switches once before reaching the satellite point, and then switches again at the satellite point. The first switch corresponds to the vHs, while the second indicates the state in which the moiré unit cell is fully occupied with four charge carriers. The coefficient of 4 represents the energetic degeneracy of the electrons' spin and valley degrees of freedom in graphene [14–16]. The area of the moiré unit cell is given by $A_m = 1/n_0 = \sqrt{3}\lambda_m^2/2 = 4/n_s$, where $n_0$ is the carrier density at which the moiré unite cell is occupied with one charge carrier, $\lambda_m$ is the moiré period (see the right image in Fig. 1(c)), and $n_s$ is the carrier density at the satellite point. The moiré period can be calculated from the observed fully filled carrier density using the following formula:

$$\lambda_m = \sqrt{\frac{8}{\sqrt{3}n_s}} \tag{5}$$

and the stacking angle ($\theta$, Fig. 1(c)) between hBN and BLG can be derived by the following equation [13]:

$$\lambda_m = \frac{(1+\delta)a_{gr}}{\sqrt{2(1+\delta)(1-\cos\theta) + \delta^2}} \tag{6}$$

where $\delta \sim 1.8\%$ is the lattice constant mismatch between graphene and hBN. From Eqs. 5 and 6 (Fig. 7(g)), the moiré parameters of D1 are estimated to be $\lambda_m \sim 13.8$ nm and $\theta \sim 0.15°$. It is known that the moiré period derived from Eq. 3 is in good agreement with that observed using scanning probe techniques [14–16,88].

For the dual-gated hBN/BLG/hBN moiré devices, Zheng *et al*. [23] and Niu *et al*. [25] reported switchable electronic ferroelectricity, where the electric field from a specific gate is anomalously screened (Fig. 8(a)). Further study [24] suggested that the layer-asymmetric moiré potential leads to localized and itinerant carrier systems that are layer-polarized (Fig. 8(b)). The localized carriers can be unidirectionally injected into the itinerant system by applying a gate voltage, a phenomenon referred to as the "electronic ratchet effect" (Figs. 8(c,d)). This effect was observed in the hBN/BLG/hBN device, where the stacking angle between the BLG and one of the hBN layers is ~0° while the other is ~15° or ~30°, This finding indicates that a significant misalignment between the top and bottom hBN layers is required to achieve a sufficient layer-asymmetric potential (Figs. 8(e,f)) [103]. These results highlight the importance of the relative stacking angle of both the top and bottom hBN layers with respect to the BLG in influencing the electronic properties. It is important to note that the dual-gated structure is essential for detecting the layer-asymmetric potential. Surprisingly, the electronic ferroelectricity and ratchet effect of hBN/BLG/hBN moiré systems remain at room temperature. Utilizing these functionalities, Yan *et al*. [26] demonstrated the operation of a low-power synaptic transistor at room temperature, indicating that the hBN/BLG/hBN moiré superlattice is a promising system for novel non-volatile memory and neuromorphic devices.

### 4-2. Tunable energy gap by displacement fields

Next, we show that the gap at the CNP opens without the displacement field and is tunable by the displacement field in the dual-gated hBN/BLG moiré devices [21]. The longitudinal resistivity at the CNP increases with increasing $|D|$, as shown in Fig. 7(e), reflecting the characteristics of BLG; the energy gap opens at the CNP with the application of the displacement field [19,20]. For the BLG devices, the energy gap ($\Delta$) can be estimated by fitting an Arrhenius plot to the thermal activation model in the high-$T$ regime, i.e., $1/\rho_{xx} \sim \exp(-\Delta/2k_BT)$, where $k_B$ is the Boltzmann constant. In contrast, variable-range hopping (VRH) [104,105] or a combination of VRH and nearest-neighbor hopping [106,107] is considered the dominant conduction mechanism for the low-$T$ regime in disordered samples. As shown in Fig. 9(a), the energy gap of D1 estimated by the activation model increases with $\Delta$[meV] ~ 110 × $|D|$[mV/nm] for positive $D$ and ~ 103 × $|D|$[mV/nm] for negative $D$, demonstrating the tunable energy gap in the hBN/BLG moiré superlattice [21]. Although the asymmetricity of the gap opening as a function of $D$—in contrast to the behavior observed for BLG misaligned with hBN—remains an open question, band-structure calculations for hBN/BLG moiré systems under an external electric field (as described in Ref. [54]) may provide valuable insight into the underlying mechanisms. The finite gap of ~1.4 meV at $D \sim 0$ can be attributed to the alignment of hBN. Recently, the tunable gap in hBN/BLG moiré superlattices was further investigated through systematic angle-controlled transport measurements using devices with a rotative top gate [55] and angle-resolved photoemission spectroscopy (ARPES) using *in-situ* gated devices (Figs. 9(d,e)) [56]. Notably, Farrar *et al*. [55] revealed that the residual displacement field ($D_r$ in Eq. 4, referred to as a crystal field) depends on the stacking angle due to the differences in the atomic configuration between BLG and hBN, which in turn affects the displacement field dependence of the gap (Figs. 9(b,c)).

The resistivity at the hole-side satellite point is insensitive to $D$ and exhibits very weak dependence on $T$, while there is almost no dependence on $D$ and $T$ at the electron side [21]. Farrar *et al*. [55] reported no thermal-activation behavior at the satellite points for the various displacement fields. Shilov *et al*. [66] and Bocarsly *et al*. [67] found the coexistence of multiple Fermi surfaces and overlapping electron-like and hole-like hBN/BLG moiré bands near the hole-side [66] and at both satellite points [67], leading to a semimetal phase. Pantaleón *et al*. [54] suggested that the low-energy main band is very sensitive to the displacement field, whereas the higher energy states around the satellite points are relatively insensitive and consist of overlapping multiple moiré bands. These results may explain the observed dependence of the satellite peaks on $T$ and $D$. The $T$-dependence away from both the CNP and the satellite points will be discussed in Section 4-5.

**4-3. Hofstadter butterfly**

In a perpendicular magnetic field, an hBN/graphene moiré superlattice shows a self-similar (i.e., fractal) quantum Hall energy spectrum known as the "Hofstadter butterfly" [14–16,63]. This phenomenon arises from the interaction between the moiré potential (a periodic electrical potential) and Landau quantization (due to magnetic fields) when their length scales are comparable. The carrier density is quantized with a unit of $n_0 = 1/A_m$ in the moiré potential, and $n_L = eB/h$ in a magnetic field. The energy spectrum is split with a unit determined by the greatest common divisor of $n_0$ and $n_L$. Figures 10(a) and (b) show the $\rho_{xx}$ and $\rho_{xy}$ maps of D1, respectively, as a function of $n$ and $B$ at $T = 1.6$ K. In the butterfly spectrum, the gapped (highly resistive) states are governed by the Diophantine equation: $n/n_0 = t\phi/\phi_0 + s$, where $\phi = BA_m$ is the magnetic flux per unit cell, $\phi_0 = h/e$ is the magnetic flux quantum, $t$ and $s$ are integers. Figure 10(c) presents the computed butterfly spectrum [53], referred to as the Wannier diagram. The butterfly spectrum exhibits a periodic feature in $1/B$ (known as Brown-Zak oscillation [108–110]), where $\phi/\phi_0 = 1/q$ (with $q$ being an integer). From its periodicity $1/B^*$, the area of the moiré unit cell can be extracted using $1/B^* = eA_m/h$, thus allowing for the determination of the moiré parameters, independent of the method described in Section 4-1. This topic, especially in the "integer" regime, is well-established [11,12,14–16,53].

Spanton *et al*. [22] revealed a topological phase with fractional values of $t$ and $s$ in the hBN/BLG moiré systems at large magnetic fields (Fig. 10(d)). This phase is classified as a fractional Chern insulator, in which the number of electrons bound to each unit cell and the net Chern number are non-integer. In the hBN/BLG moiré system, the Chern bands are formed at large magnetic fields by the electrons in the periodic moiré potential, which can be controlled by the stacking angle. Additionally, penetration field capacitance measurements can be employed to determine whether the bulk of the system is gapped. Jeong *et al*. [27] reported strongly correlated insulating states at integer fillings $n/n_0 = -1, -2$ in the hBN/BLG moiré devices through transport measurements at moderate magnetic fields (Fig. 10(e)). This behavior is similar to the correlated insulator states in the magic-angle TBG systems, which emerge at integer moiré fillings [70,111,112]. These observations [22,27] suggest that the hBN/BLG moiré system could serve as a tunable platform for exploring strongly correlated phenomena.

### 4-4. Quantum valley currents

A valley is a degree of freedom associated with electrons in several solid-state systems [57]. A material with a gapped Dirac dispersion and broken spatial-inversion symmetry can exhibit topological valley currents even in the absence of broken time-reversal symmetry [57,113,114], as described below. One example is hBN/graphene moiré superlattices which have an energy gap at the CNP without the application of a displacement field [58,59,115,116]. Such a gapped Dirac material has the accumulated Berry curvature around the band edges (Fig. 11(a) [57]), so-called "hot spot." The semiclassical equation of motion for electrons in a system with finite Berry curvature is given by [113]:

$$v(k) = \frac{\partial \varepsilon(k)}{\hbar \partial k} - \frac{e}{\hbar} E \times \Omega(k) \quad (7)$$

where $v$ is the group velocity of Bloch electrons, $\varepsilon$ is the energy, $k$ is the wavevector, $E$ is the in-plane electric field, and $\Omega$ is the Berry curvature. The second term of the right side of Eq. 7 is called the "anomalous velocity," where the Berry curvature acts as a magnetic field in momentum space, analogous to the Lorentz term in real space. In a system with both time-reversal and spatial-inversion symmetry, the time-reversal operation changes the sign of $v$, $k$, whereas leaving $E$ unchanged. Conversely, the spatial-inversion operation changes the sign of $v$, $k$, and $E$. Therefore, a time-reversal symmetric system requires $\Omega(k) = -\Omega(-k)$, whereas a spatial-inversion symmetric system requires $\Omega(k) = \Omega(-k)$. Hence, the Berry curvature vanishes in a system with both time-reversal and spatial-inversion symmetry. In gapped graphene systems, where spatial-inversion symmetry is broken ($\Omega(k) \neq \Omega(-k)$), the Berry curvature has opposite signs for each valley $K$ and $K'$ ($\Omega(k) = -\Omega(-k)$). This induces an anomalous velocity that points in opposite directions for electrons in each valley, leading to the generation of valley currents at zero magnetic field (Fig. 11(b)).

The detection of valley currents through electrical transport is challenging due to their charge-neutral characteristics. To address this, a nonlocal transport configuration has been widely employed to generate and detect valley currents, as shown in Fig. 11(c). In an H-shaped geometry, which is often part of a Hall bar, a transverse valley current is generated by an electric current $I$ between the left-side terminals. This occurs due to the anomalous velocity of electrons in each valley, a phenomenon is called the valley Hall effect. The valley current is then converted into a voltage drop ($V_{nl}$) between the right-side terminals, referred to as the inverse valley Hall effect. This approach has been demonstrated in various systems, including hBN/SLG moiré superlattices [115,116], dual-gated BLG devices [35–37], hBN/BLG moiré superlattices [25,58,59], and odd-layer $MoS_2$ devices [117]. The nonlocal resistance, defined as $R_{nl} = V_{nl}/I$, is used to characterize the valley (flavor) currents [118,119]. Fig. 11(d) shows $R_{nl}$ of the hBN/BLG moiré device as a function of $n$ at $T = 6$ K [58]. The nonlocal resistance exhibits a peak at the CNP, where Berry curvature is accumulated [57]. In the regime where $\sigma_{xx} \gg \sigma_{xy}^v$ (with $\sigma_{xx} = 1/\rho_{xx}$ representing the longitudinal local conductivity and $\sigma_{xy}^v$ denoting the valley Hall conductivity), the nonlocal resistance can be expressed using the bulk-mediated diffusion model [115,118]:

$$R_{nl} = \frac{W}{2l_v}(\sigma_{xy}^v)^2 \rho_{xx}^3 \exp\left(-\frac{L}{l_v}\right) \tag{8}$$

where $L$ is the channel length, $W$ the channel width, and $l_v$ the intervalley scattering length. Consequently, the nonlocal resistance is expected to scale cubically with the local resistivity ($R_{nl} \sim \rho_{xx}^3$, as indicated by the dashed line in Fig. 11(e)) [25,35–37,58,59,114–117]. This leads to the thermal activation energy of the nonlocal transport being three times higher than that of local transport [114]. Arrighi et al. [59] reported that the scaling relation between nonlocal resistance and local resistivity in the hBN/BLG heterostructure varies for stacking angles of 0°, 30°, and 60° (Fig. 11(f)). The relation for the valley Hall effect ($R_{nl} \sim \rho_{xx}^3$) is only observed when the stacking angle is 0°, even though the hBN/BLG moiré pattern is expected to exhibit the same structure at 60°. Numerical simulations suggest that in-plane atomic structural relaxation within the moiré unit cell occurs, leading to differences in the band structures at 0° and 60° [59]. The intervalley scattering length can be estimated from the channel length dependence of $R_{nl}$ using the multi-terminal devices. The typical extracted values of $l_v$ are ~1–2.5 μm for graphene-based systems [35–37,59,115] and ~1 μm for the $MoS_2$ systems [117]. The valley current is reduced by intervalley scattering caused by atomic-scale defects. Such defects are rare in the graphene-based systems; therefore, the valley bulk current is most likely limited by the device edges.

A key issue in this topic is the concept of the "quantum limit," which is considered to hold in the insulating regime ($\sigma_{xx} \ll \sigma_{xy}^v$, i.e., the valley Hall angle $\sigma_{xy}^v/\sigma_{xx}$ above ~1) and the emergence of the quantum valley Hall state, which is associated with a non-local resistance of the order $\sim h/e^2$ [37,116,120]. In the hBN/SLG moiré device, the nonlocal resistance at the hole-side satellite point was found to be higher than the local resistivity in the low-$T$ regime [116]. In this regime, the nonlocal resistance at the hole-side satellite point is within the quantum limit, i.e., $R_{nl} \sim h/2e^2$, and a cubic scaling between $R_{nl}$ and $\rho_{xx}$ fails. This signature was also observed in the non-aligned BLG device (D2) in the low-$T$ regime (Fig. 11(e)) [37]. The mechanism was considered to be edge state transport in the quantum limit, rather than a bulk-related interpretation [116]. However, understanding the mechanism behind these results calls for further theoretical and experimental investigation of the valley current in the insulating regime.

We note that, although the nonlocal signal associated with valley current has been widely studied, its origin remains a topic of debate. It has been suggested that a nonlocal signal can also arise from non-topological edge currents [121], locally opened energy gaps [122], charge accumulation at the edges [123], and the orbital moment-induced Hall effect (orbital Hall effect) [124,125]. On the other hand, the detection of the valley current signals has been demonstrated through optical pumping with circularly polarized light [126,127], spatial mapping of Kerr rotation [128], and transport through the "kink state" that emerges in a one-dimensional BLG channel between two insulating regions with opposite sign energy gaps [129–131].

### 4-5. Flavor symmetry breaking

The flavor symmetry in BLG and hBN/BLG systems at the CNP is sensitive to displacement and magnetic fields. Figure 12(a) presents the $\rho_{xx}$ map of D1 as a function of $D$ and $B$ at $n = 0$ (CNP). This phase diagram reveals two "conductance spikes (or resistance dips)" indicated by the white lines, where the energy gap is abruptly closed (Fig. 12(b)). This behavior suggests first-order phase transitions accompanied by flavor-symmetry breaking [21]. The central insulating region enclosed by the white lines is associated with the canted antiferromagnetic phase (Fig. 12(c)), in which electrons in different valleys (or layers) exhibit opposite in-plane spin orientations, with a slight out-of-plane spin orientation to minimize the antiferromagnetic exchange and Zeeman energy [132]. The insulating regions at high displacement fields are attributed to layer-(valley-)polarized phases (Fig. 12(d)) [133]. The phase diagrams at the CNP of BLG, both with and without alignment to hBN, appear almost similar in experimental observations [21,133], warranting further investigation for the latter case.

Lastly, let us discuss the properties of hBN/BLG moiré devices away from the CNP and the satellite points. Figure 13(a) shows the $\rho_{xx}$-$n$ curve of D1 for various $T$. As shown in Fig. 13(b), the resistance-temperature ($R$-$T$) characteristics in D1 for carrier densities between the CNP and the satellite points exhibit metallic behavior with a strong $T$-dependence, following a power-law relationship of the form $\sim T^p$ with $p \sim 1$–$2$, depending on $n$. It is well established that the $R$-$T$ characteristics in the doped regimes ($n \neq 0$) are basically due to phonon contributions in the graphene family, leading to $p \sim 1$ [134]. In contrast, electron-electron interaction—specifically, "Umklapp effects," where a pair of electrons scatters due to Coulomb repulsion and transfers momentum to the lattice—can dominate the scattering process in hBN/graphene moiré superlattices [64–66,135], with predictions suggesting $p \sim 2$ [64,135–137]. The Umklapp effect can be toggled on/off due to the controllable Fermi wavevector via the carrier density, and reciprocal moiré lattice vector, which is affected by the stacking angle (see Fig. 13(c) and the inset of Fig. 13(b)). Recent studies by Jat *et al*. and Shilov *et al*. have demonstrated modulation of $p \sim 1$–$2$ through the carrier density [65,66], stacking angle, and displacement field [65]. Notably, the THz excitation studied in [66] can increase the electron temperature while keeping the lattice cold, allowing for a clearer distinction between contributions from electron-phonon interactions and electron-electron interactions. Contrary to the findings in ref. [65], the power-law behavior ($p \sim 1$–$2$) was also observed in our device D2 [37], despite its stacking angle between hBN and BLG not being close to 0° (we did not observe the satellite points within the accessible gate voltage range). A possible explanation for this observation is the influence of the moiré pattern with $\lambda_m \sim 6.3$ nm ($\theta \sim 2°$, estimated from its optical image [37]) or the modulation of the band structure and/or layer polarization due to the asymmetric relative stacking angle of both the top and bottom hBN with respect to the BLG [103]. A more detailed assessment of the scattering mechanisms in hBN/BLG heterostructures and their dependence on stacking angles remains an open question.

On the other hand, significant departures from such an assignment are observed in D1 at energy states lower than the hole-side satellite point ($n \sim -3.0 \times 10^{12}$ cm$^{-2}$ and $D \sim 35$ mV/nm, see Fig. 13(d)) [21]. This state exhibits insulating behavior (combined with a "resistance bump") in the low-$T$ limit, contrasting with the conventional phonon contribution. In another hBN/BLG moiré device shown in Fig. 13(e), we observed a resistance bump accompanied by a nearby drop in the low-$T$ limit in the regions between the hole-side satellite point and the CNP. Although this type of insulating behavior has also been observed in other graphene systems, its origin has not yet been definitively identified, and it has been routinely referred to as a correlated insulator or Mott insulator in several literatures [7,9]. "Hidden orders," such as (unconventional) charge/spin/valley density waves, are potential candidates for the origin of these departures [138,139], but a more careful investigation is required as future work. Additionally, other physics, such as "fractional" phenomena, may compete with these orders. More generally, phenomena characterized by a resistance bump (and/or a nearby drop) can arise from various broken symmetries or topological orders due to many-body effects [140–145]. Such orders are always competing within the narrow energy band associated with vHs, and there are several approaches to fix the survivor and "resolve the singularities."

## 5. Summary and perspective

We have discussed the transport properties in hBN/BLG moiré superlattices, where both the energy band structure and the carrier concentration can be tuned by dual gating. Utilizing state-of-the-art fabrication and measurement techniques, a variety of intriguing phenomena have been observed [21–27,58,59,65–67,102,146], and this field has been rapidly growing. The hBN/BLG moiré systems hold great potential, including rich phase diagrams across numerous parameter spaces, as evidenced by recent studies on dual-gated BLG [28–37] and rhombohedral multilayer graphene devices [38–41], as well as unique degrees of freedom, such as the relative stacking angle between the top and bottom hBN layers [23,103]. We note that both aligned and non-aligned BLG exhibit tunability in the gap opening at the CNP, but in the hBN/BLG moiré case additionally the satellite points of the moiré-induced minibands in both the conduction and valence regimes significantly narrow the bandwidth of the main bands and give rise to additional vHs. Moreover, the properties of BLG can be modified not only by hBN but also by other adjacent layers; for instance, spin-orbit coupling is enhanced when a transition metal dichalcogenide layer is in proximity to BLG, such as $WSe_2$ [147] or $MoS_2$ [148]. It has also been experimentally demonstrated [149–151] and theoretically predicted [152,153] that the moiré/periodic potential in adjacent twisted stacks (such as twisted double bilayer $WSe_2$ [149] or twisted hBN [150,152]) or patterned gates [151,153] can modify the BLG band structure and topology. Furthermore, the band structure of BLG can be engineered by aligning both the top and bottom hBN layers with the encapsulated BLG [154]. These concepts could be combined with the hBN/BLG moiré system for further band engineering.

The concept of "superlattice" can be traced back to the celebrated works by Esaki and Tsu [155] and the high-electron-mobility transistor developed by Mimura *et al*. [156], which serves as a platform for quantum Hall and mesoscopic physics. The ability to control the energy gap with an external electric field suggests that arbitrary structures can be defined electrostatically. Therefore, based on hBN/BLG moiré superlattices, advancements in nanofabrication techniques should enable the construction of quantum devices, such as quantum dots (or single-electron transistors: SETs) [102, 157], quantum point contacts (QPCs) [158], quantum ring interferometers [159], and more, including their "aufheben." It is important to note that the QPC is a basic building block for "fermionic quantum optics" [160], which contrasts with the "single-electronics" in the quantum-dot/SET [157]. Although the QPCs have been realized in SLG [161,162] and both the QPCs and quantum dots have been explored in BLG [163,164]—even in the absence of near-zero–degree alignment with hBN—investigations of such devices within the specific context of hBN/BLG moiré superlattices remain comparatively limited [102]. Consequently, substantial opportunities remain for future research, including, for example, the electrical control of transitions between competing orders, particularly those of topological character. In this regard, low-dimensional and nanoscale physics should open a door for the next level of research combined with carbon physics [1–4,165,166]. This review should lay the foundation for the "global phase diagram" of hBN/BLG moiré superlattices and beyond, including the fabrication and study of nanostructures like quantum dots/SETs, QPCs, Josephson Junctions, and their integrated structures in quantum circuits.

**Acknowledgements**

We thank all our collaborators, although we do not list them up all. In particular, T. Taniguchi and K. Watanabe played a crucial role for providing us ultra-pure hBN crystals. We also thank H. Osato, E. Watanabe, and D. Tsuya from the NIMS Nanofabrication Facility. This work was partially supported by JPSJ KAKENHI Grant No. 21H01400, and the World Premier International Research Center

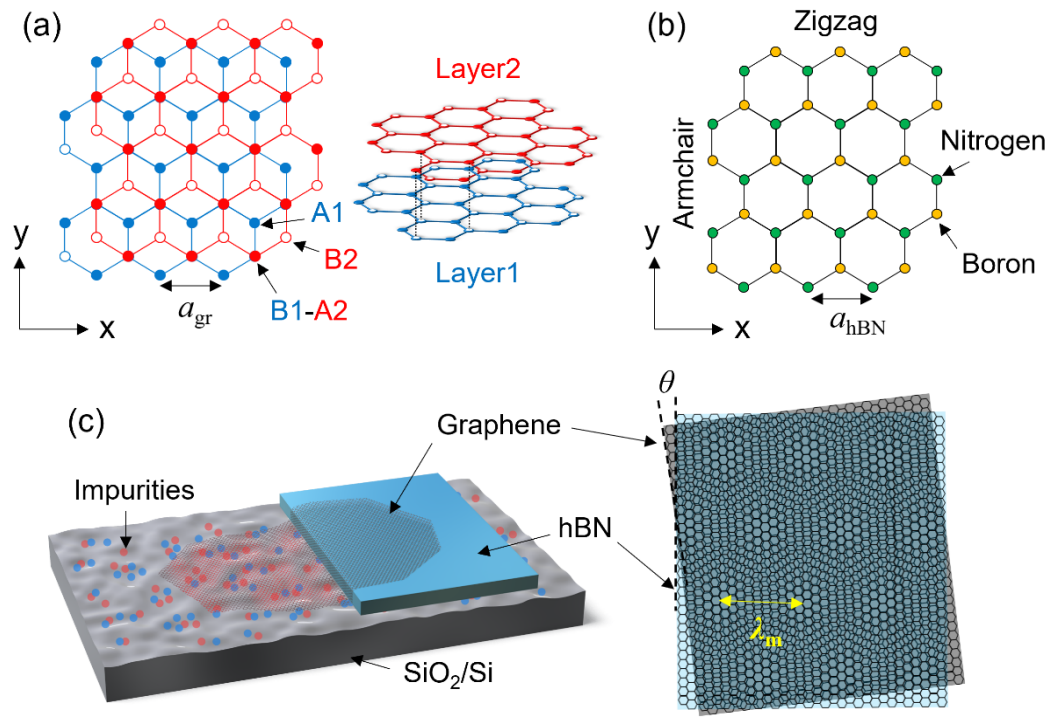

**Fig. 1. Schematic structure.** (a) Bernal bilayer graphene lattice. $a_{gr}$ is the lattice constant of graphene. The filled and unfilled markers represent the atoms in the *A*- and *B*-sublattices, respectively. (b) hBN lattice. $a_{hBN}$ is the lattice constant of hBN. (c) Left: Graphene on a SiO$_2$/Si substrate and hBN. Right: hBN/graphene moiré superlattice.

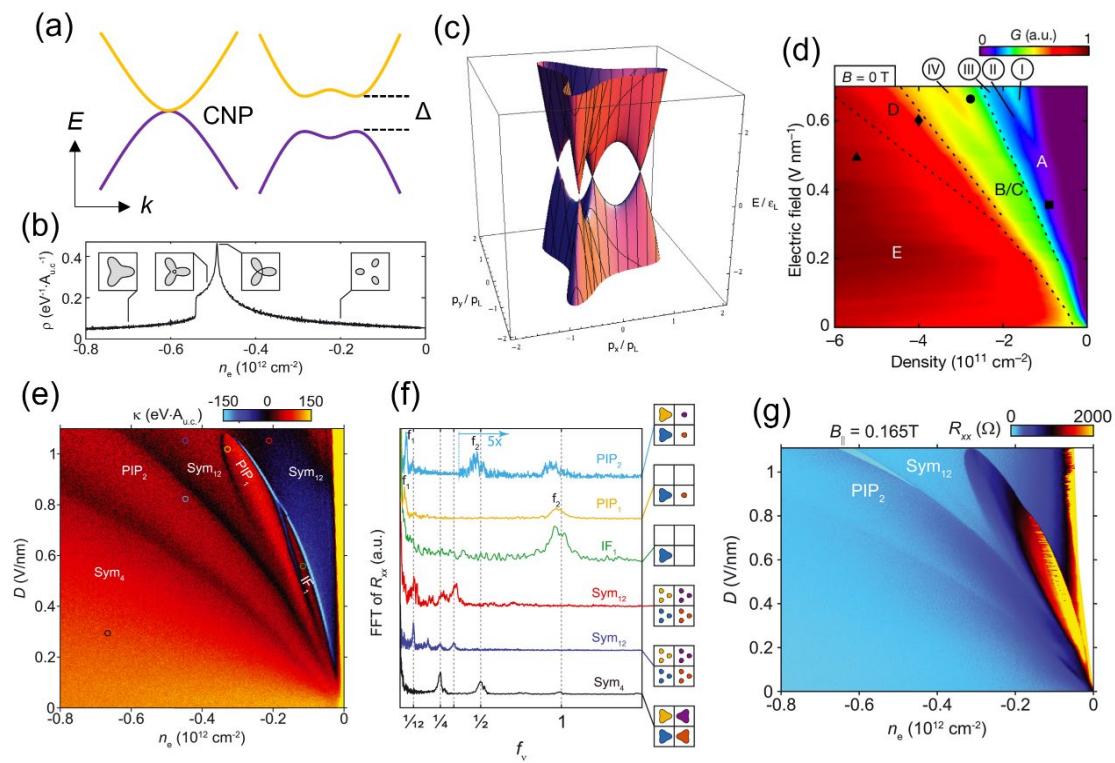

**Fig. 2. Electronic structure of BLG.** (a) Band structure of pristine BLG (left) and that with broken inversion symmetry (right). (b) Density of states of BLG as a function of carrier density with the interlayer potential difference. Insets: Fermi contours at the indicated carrier densities. (c) Three-dimensional plot of the low-energy dispersion of BLG near the $K$ point, highlighting the trigonal warping effect. (d) Conductance map of the BLG device as a function of electric field and carrier density, exhibiting Stoner ferromagnetic phases labelled A–E. (e-g) Flavor symmetry breaking in BLG. (e) Inverse compressibility measurements, and (f) the fast Fourier transform (FFT) of the Shubnikov-de Haas oscillations at the points indicated in (e), show different Fermi pocket occupations associated with flavor symmetry breaking. (g) In a small in-plane magnetic field, the superconducting phase (the bright cyan region) emerges between the two phases. (b,e-g) Reproduced with permission from [34], AAAS. (c) Reproduced with permission from [52], IOP publishing. (d) Reproduced with permission from [33], Springer Nature.

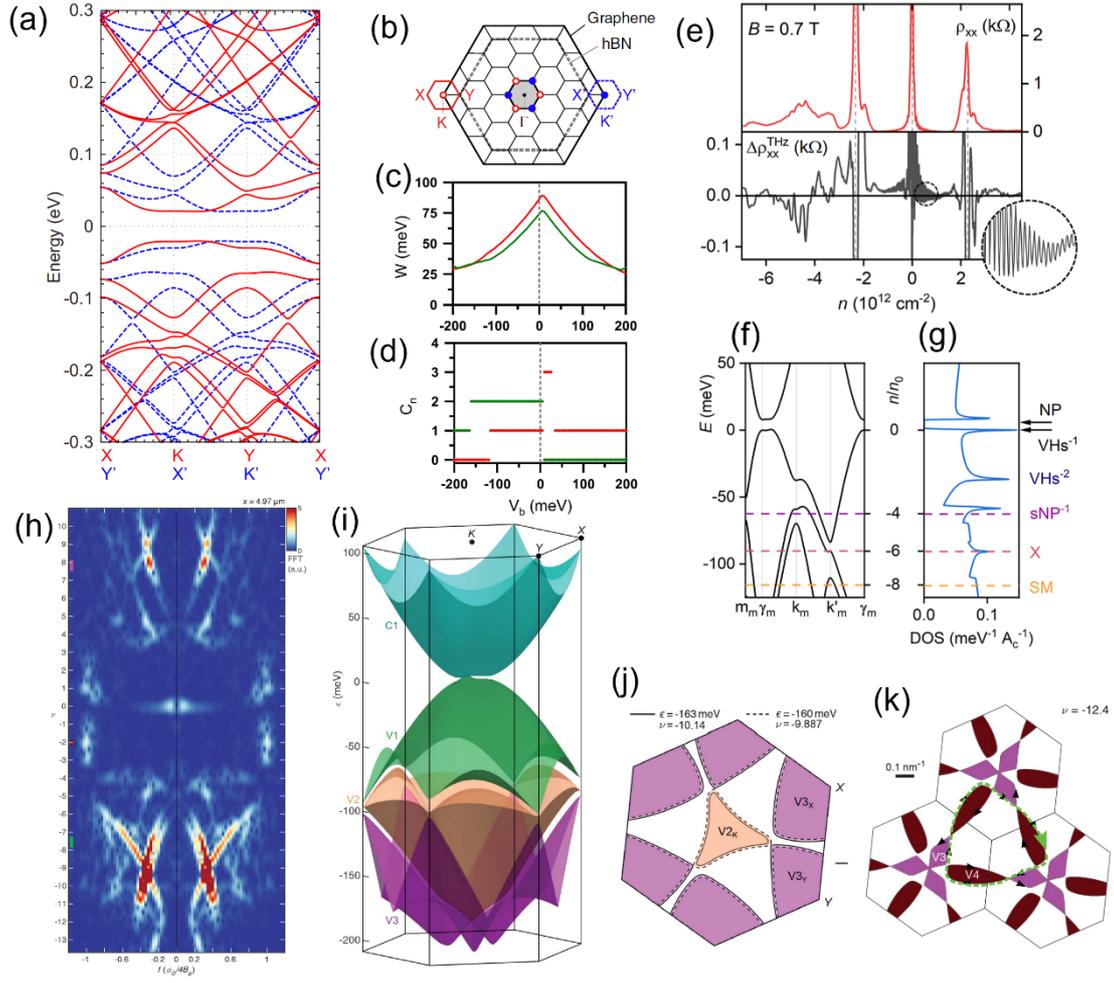

**Fig. 3. Electronic structure of hBN/BLG moiré superlattices.** (a) Band structure of the hBN/BLG moiré superlattice at a stacking angle of 0°, calculated using the continuum model (Eq. 1). The *k*-space path is shown in (b). (b) BZ folding in the hBN/graphene moiré superlattice. (c) Bandwidth and (d) the Chern number of the middle band of the hBN/BLG moiré superlattice as a function of the displacement field. (e-g) Investigation using THz excitation. (e) Carrier density dependence of the longitudinal resistivity (top) and THz photoresistivity (bottom) of the hBN/BLG moiré device. The THz signal exhibits more detailed features and aligns with (f) the calculated band structure and (g) the density of states. (h-k) Characterization through de Haas-van Alphen spectroscopy. (h) FFT amplitude map of the de Haas-van Alphen oscillations in the hBN/BLG moiré device as a function of the moiré filling factor and the FFT frequency. The multiple frequencies at higher moiré fillings indicate the presence of narrow moiré bands with overlapping Fermi surfaces. (i) Calculated band structure, exhibiting the middle valence band (V1) partially overlapped with two remote valence bands (V2 and V3). (j) Multiple Fermi surfaces at a filling of −10.14. (k) Band structure at a filling of −12.28. The green trajectory depicts the shortest magnetic coherent breakdown orbit. (a,b) Reproduced with permission from [53], Copyright (2014) American Physical Society. (c,d) Reproduced with permission from [54], IOP publishing. (e-g) Reproduced with permission from [66], Copyright (2024) American Chemical Society. (h-k) Reproduced with permission from [67], AAAS.

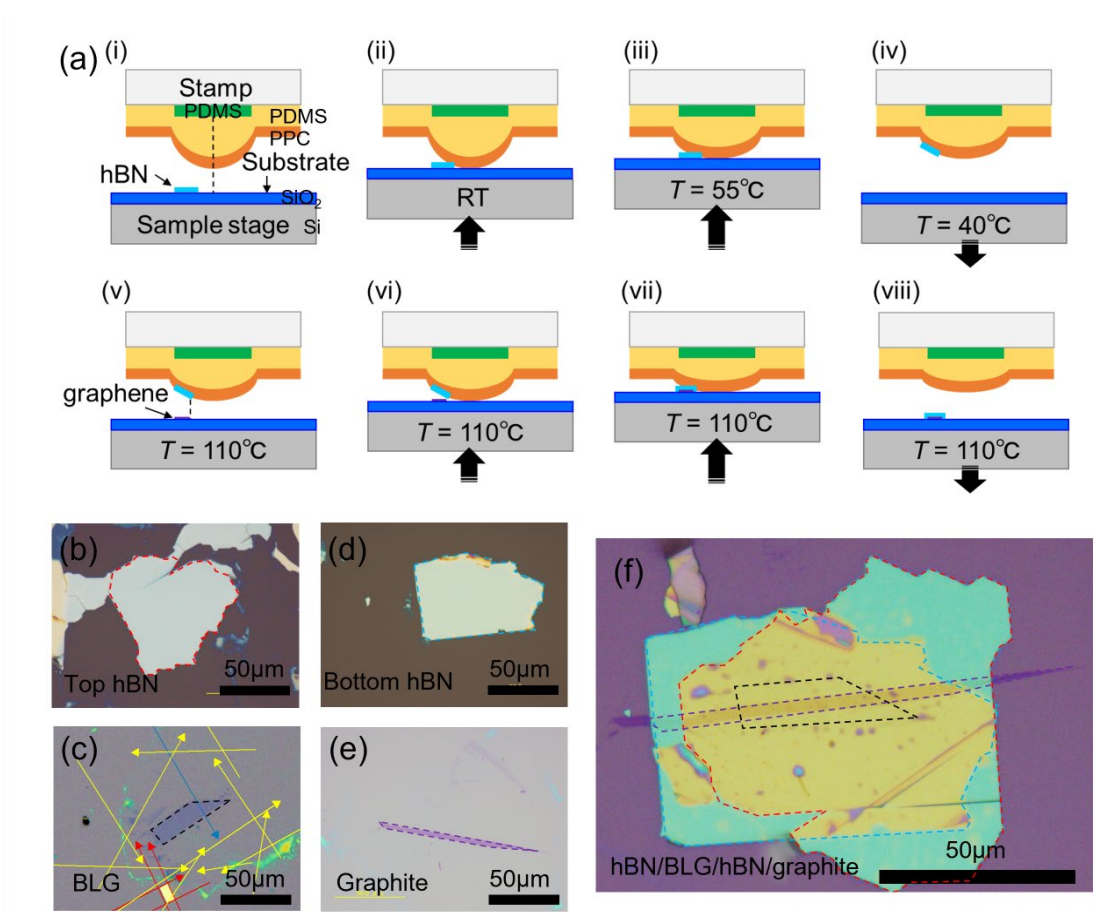

**Fig. 4. Transfer process.** (a) Schematic illustration of the transfer process flow. (b-f) Optical images of each flake before transfer and the completed hBN/BLG/hBN/graphite stack. (a) is modified and reproduced from [84].

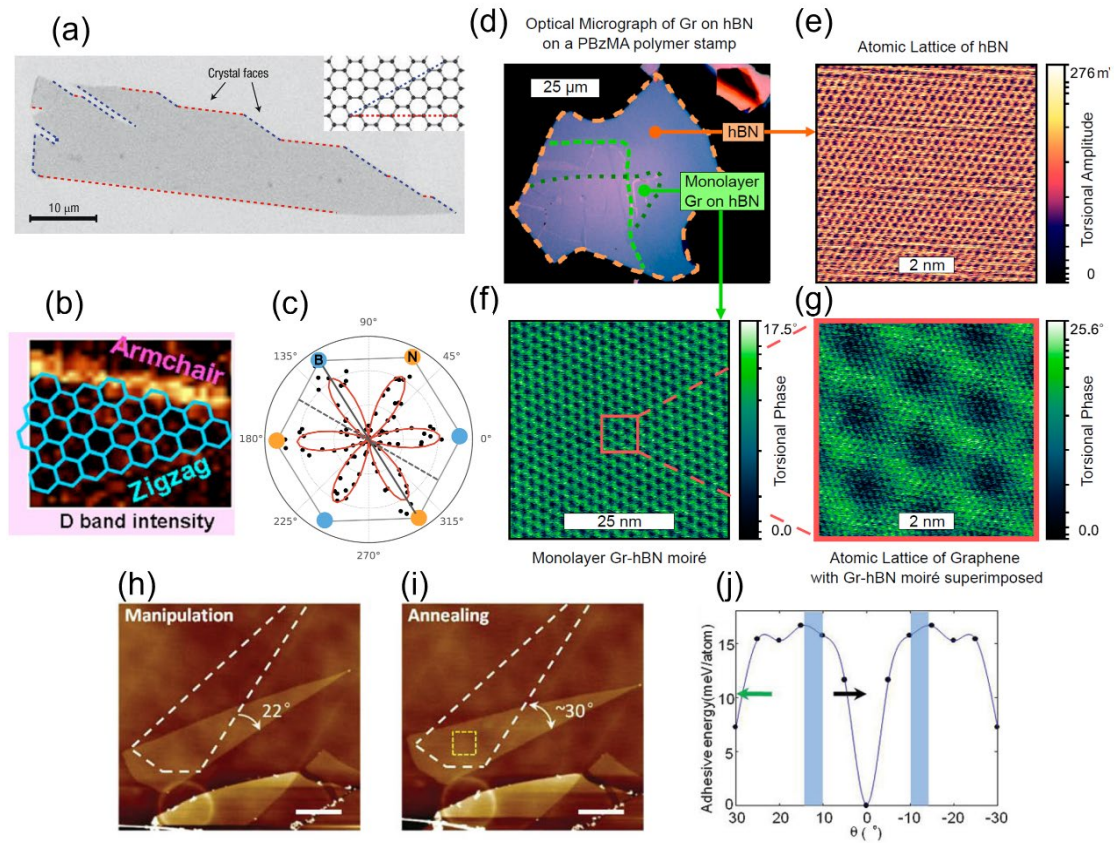

**Fig. 5. Alignment of the stacking angle.** (a) Scanning electron micrograph of a graphene flake. The zigzag and armchair edges (schematically shown in the inset) tend to appear along the long straight edge. (b) Raman mapping near the graphene edge. The intensity of the D band in the Raman spectrum is sensitive to the edge chirality and is pronounced at the armchair edge. (c) Intensity of the SHG signal as a function of laser polarization, showing the minimum (maximum) values when the polarization is parallel to the zigzag (armchair) edge. (d-g) Visualization by TFM. (d) Optical image of a graphene/hBN heterostructure on a polymer stamp. (e) TFM image of the hBN atomic lattice measured at the position marked in (d). (f) TFM image of the graphene/hBN moiré pattern at the position marked in (d). (g) Zoom image of (f), exhibiting both the atomic graphene lattice and moiré pattern simultaneously. (h-j) Thermally induced rotation of graphene on hBN. (h) AFM image of graphene on hBN. The graphene is mechanically rotated to ~22° from the initial angle of 30° (resulting in the stacking angle of $\theta \sim 8°$ between graphene and hBN). (i) Graphene is rotated to 30° (i.e., $\theta \sim 0°$) measured from the initial angle after annealing. (j) Simulated adhesive energy of graphene/hBN stacking as a function of stacking angles. The thermally induced rotation tends toward $\theta \sim 0°$ (30°) when the angle before annealing is less than (greater than) ~|12|°. (a) Reproduced with permission from [4], Springer Nature. (b) Reproduced with permission from [86], AIP Publishing. (c) Reproduced with permission from [88], National Academy of Sciences. (d-g) Reproduced with permission from

[89], National Academy of Sciences. (h-j) Reproduced with permission from [91], Copyright (2016) American Physical Society.

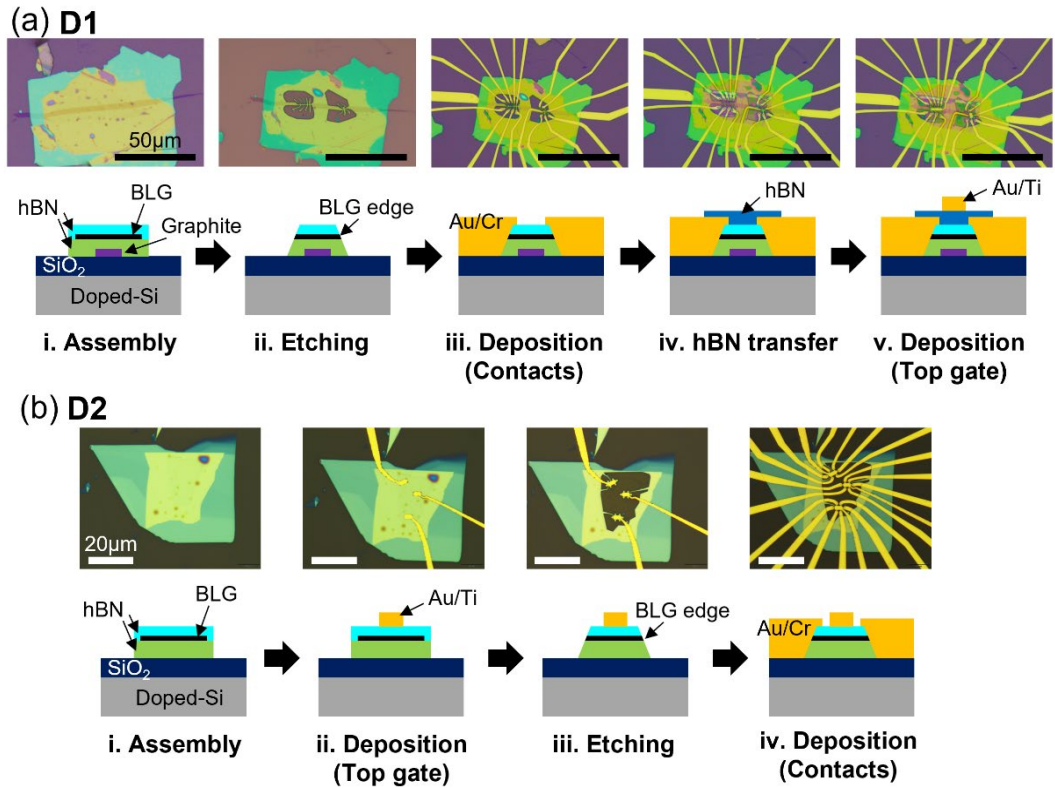

**Fig. 6. Fabrication process.** (a) Schematic flow of the fabrication process (bottom) and corresponding optical images (top) of device D1. (b) The same for device D2.

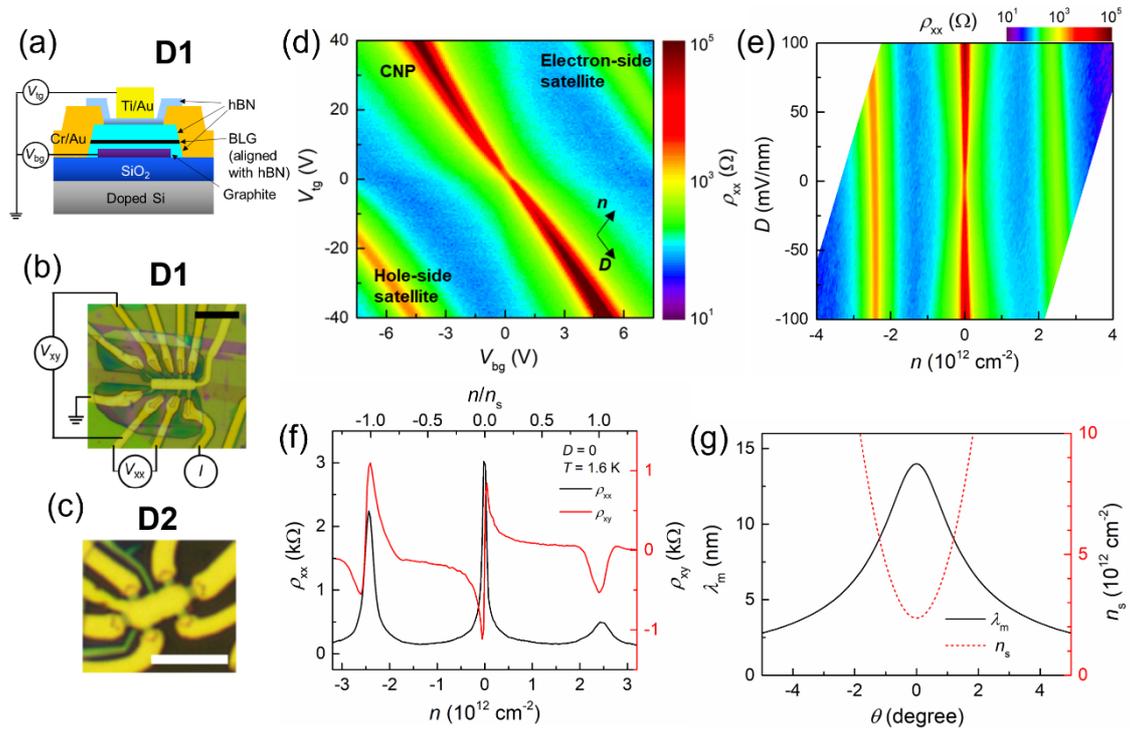

**Fig. 7. Dual-gated transport characteristics.** (a) Schematic cross-section of device D1. (b) Optical image of D1 and (c) D2. The scale bars in (b) and (c) correspond to 10 μm and 5 μm, respectively. (d) Longitudinal resistivity map as a function of top-gate and back-gate voltages for D1 measured at a temperature of 1.6 K. (e) Longitudinal resistivity map in the carrier density versus displacement field plane, converted from (d). (f) Longitudinal (black) and Hall resistivity (red) as a function of carrier density. (g) Moiré period (Eq. 6) and carrier density corresponding to the satellite points (Eq. 5) as a function of stacking angle. (a,b) Reproduced with permission from [21], Copyright (2022) American Physical Society. (c) Reproduced with permission from [37], Copyright (2024) American Physical Society. (d) is modified and reproduced from [21].

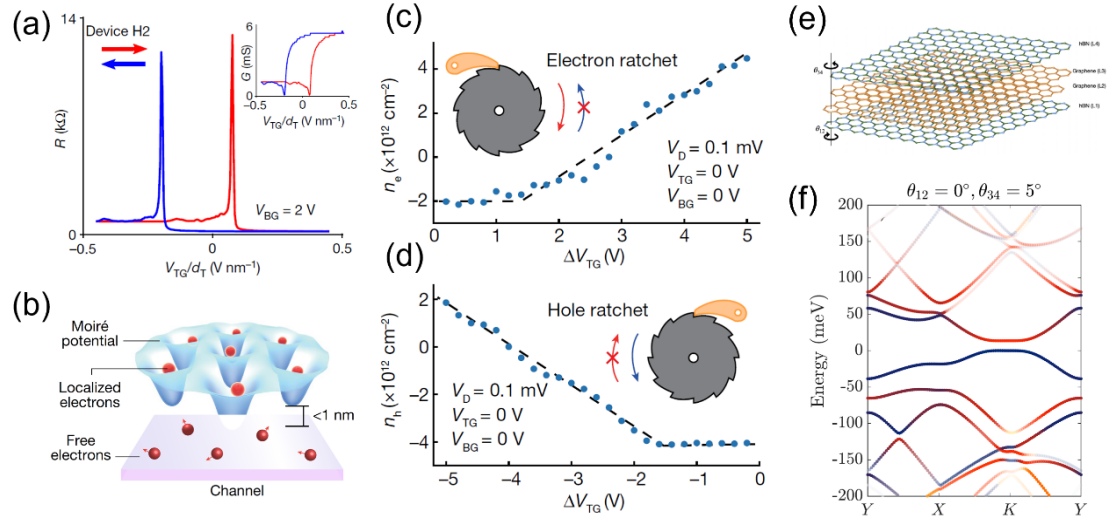

**Fig. 8. Influence of the relative stacking angle between top and bottom hBN.** (a) Hysteric behavior in the hBN/BLG/hBN moiré device. The angle between the top (bottom) hBN and the BLG is ~0° (~30°). The position of the resistance peak substantially shifts depending on the scan direction of the displacement field, leading to significant hysteresis. (b-d) Electronic ratchet effect. (b) Asymmetric potential splits the system into localized and itinerant charges. The moiré potential at the top hBN/BLG interface traps charges, while the bottom layer serves as the lateral charge transport channel. (c-d) Injected carrier density as a function of the top-gate voltage scan range. The top-gate sweep allows (c) electrons or (d) holes to be injected from the localized system into the itinerant one, but prevents them from returning to the localized system during the reverse sweep. (e-f) Layer asymmetric potential-induced layer polarization. (e) Schematic of hBN-encapsulated BLG with two twist angles. (f) Calculated band structure of the hBN/BLG/hBN system for the angles between top (bottom) hBN and BLG of 0° (5°). The color represents the projection of the wave-function weights for each graphene layer. (a) Reproduced with permission from [23], Springer Nature. (b-d) Reproduced with permission from [26], Springer Nature. (e,f) Reproduced with permission from [103], Copyright (2022) American Physical Society.

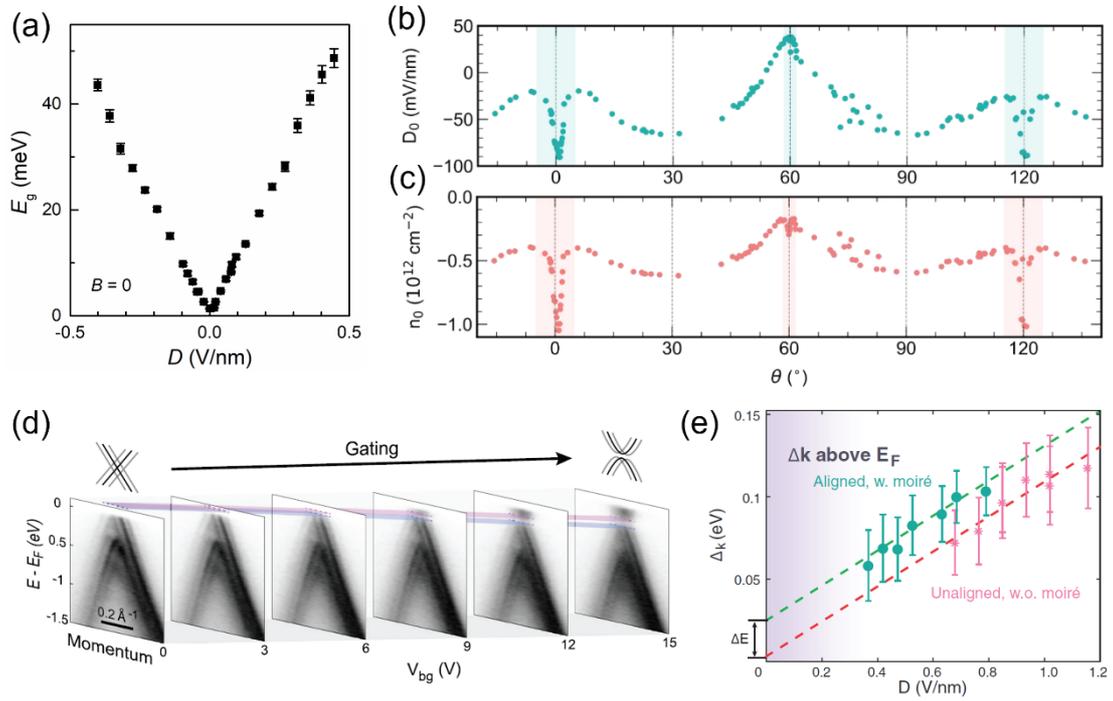

**Fig. 9. Tunable energy gap.** (a) Energy gap of device D1 as a function of the displacement field. (b,c) Systematic angle-controlled transport measurements. (b) Crystal field and (c) the residual carrier density as a function of the stacking angle between hBN and BLG, reflecting the 120° symmetry. (d-e) ARPES measurements for the *in-situ* gated BLG device. (d) Evolution of the band dispersion of the BLG/hBN moiré device with varying the gate voltages. As the gate voltage increases, the energy gap opens at the CNP. (e) Energy gap extracted from the ARPES spectra for the BLG with (green) and without (pink) alignment to hBN, as a function of the displacement field. (a) Reproduced with permission from [21], Copyright (2022) American Physical Society. (b,c) Reproduced with permission from [55], Copyright (2025) American Chemical Society. (d,e) Reproduced from [56], Wiley-VCH.

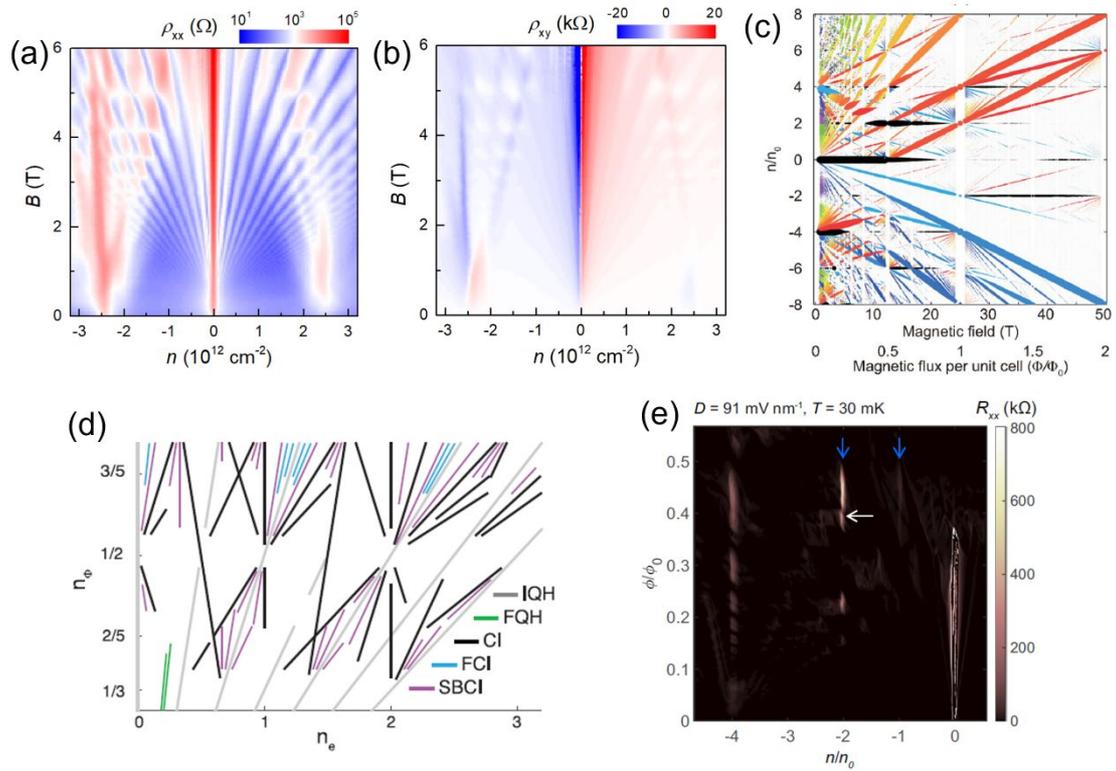

**Fig. 10. Spectra in a perpendicular magnetic field.** (a) Longitudinal resistivity and (b) Hall resistivity map of device D1 as a function of carrier density and magnetic field, measured at a temperature of 1.6 K. (c) Wannier diagram of the hBN/BLG moiré superlattice. The thickness of the lines is proportional to the gap width, and the color indicates the difference in the quantized Hall conductivity. (d) Magnetocapacitance measurement of the hBN/BLG moiré device with the graphite dual gate shows several gap spectra distinguished by color: IQH: integer quantum Hall ($s = 0$, $t \in \mathbb{Z}$), FQH: fractional quantum Hall ($s = 0$, $t$: fractional), CI: Hofstadter Chern insulators ($s, t \in \mathbb{Z}$, $s \neq 0$), SBCI: symmetry-broken Chern insulators ($s$: fractional, $t \in \mathbb{Z}$), and FCI: fractional Chern insulators ($s, t$: fractional). (e) Magnetotransport of the graphite dual-gated hBN/BLG moiré device exhibits the correlated insulating states observed at the moiré filling $n/n_0 = -1$, and $-2$ indicated by the blue arrows. (d) Reproduced with permission from [22], AAAS. (e) Reproduced from [27], Springer Nature.

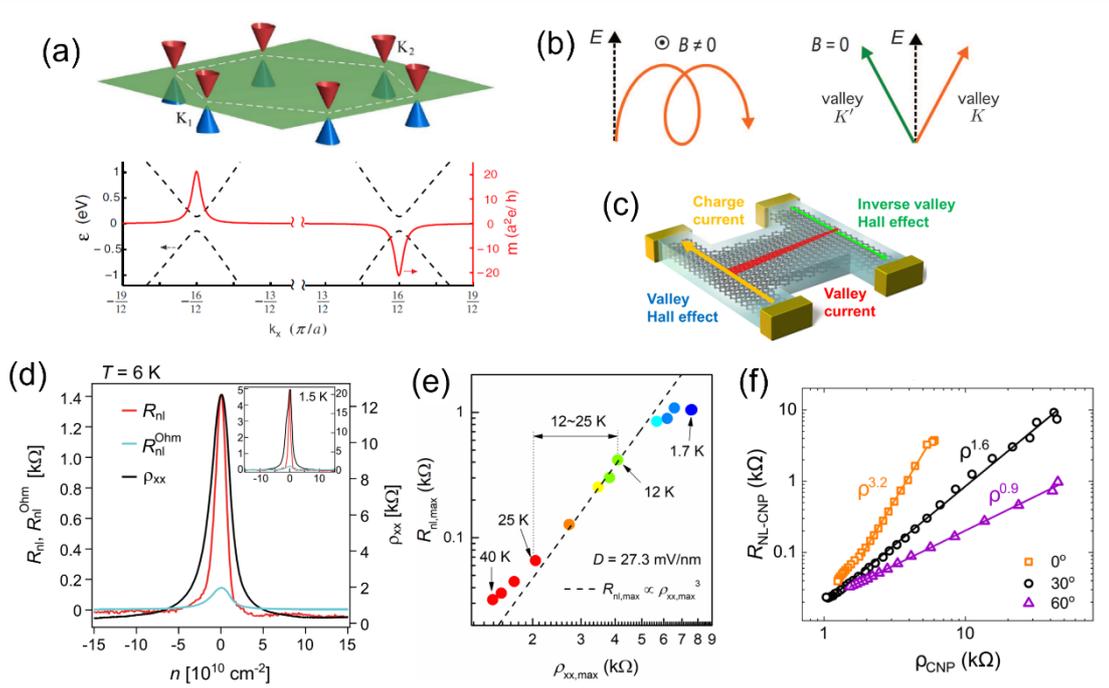

**Fig. 11. Nonlocal transport.** (a) Energy bands with an energy gap (top panel) and the orbital magnetic moment of the conduction bands (solid lines at the bottom panel) of graphene with broken inversion symmetry. The Berry curvature exhibits a distribution similar to that of the orbital magnetic moment. (b) Left: Schematic of drifting cyclotron orbits in magnetic fields. Right: Schematic of skewed motion induced by the Berry curvature. The direction has opposite for each valley. $E$ represents an electric field. (c) Schematic of valley current generation and detection in the nonlocal transport configuration. (d) Nonlocal resistance ($R_{nl}$), the Ohmic contribution ($R_{nl}^{Ohm}$), and the local resistivity ($\rho_{xx}$) of the hBN/BLG moiré device as a function of the carrier density, measured at a temperature of 6 K. Inset: The same measurements at a temperature of 1.6 K. (e) Scaling analysis of device D2 for various temperatures. The dashed line represents a fit to the cubic relation ($R_{nl} \sim \rho_{xx}^3$). (f) Scaling relation between local resistivity and nonlocal resistance of the hBN/BLG moiré devices with stacking angles of 0°, 30°, and 60°. The difference in power-law dependence for each angle alignment indicates the non-identical states even between the 0° and 60° configurations. (a) Reproduced with permission from [57], Copyright (2007) American Physical Society. (b) Reproduced with permission from [115], AAAS. (c,d) Reproduced with permission from [58], AIP Publishing. (e) Reproduced with permission from [37], Copyright (2024) American Physical Society. (f) Reproduced from [59], Springer Nature.

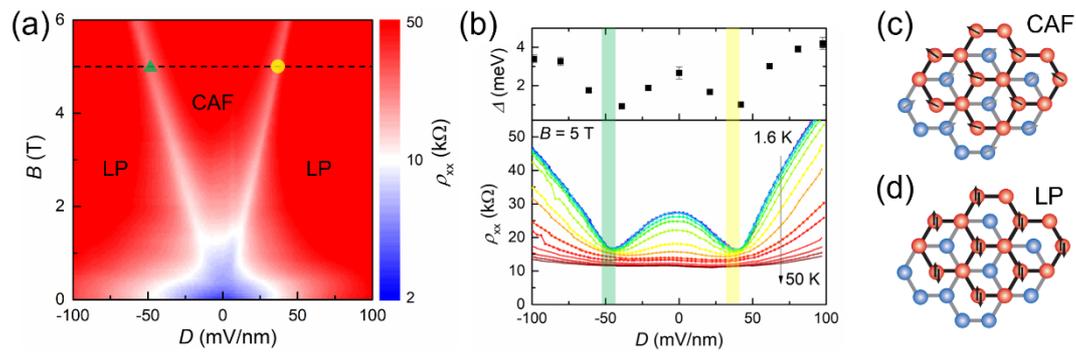

**Fig. 12. Phase transition at the CNP.** (a) Mapping plot of the longitudinal resistivity of device D1 as a function of displacement field and magnetic field at the CNP. The white lines represent phase transitions. LP: layer-polarized state. CAF: canted-antiferromagnetic state. (b) Energy gap and longitudinal resistivity of D1 as a function of displacement field at a magnetic field of 5 T, corresponding to the horizontal dashed line shown in (a). The colored stripes correspond to the marked points in (a). (c) Schematic of the CAF and (d) LP in BLG. (a,b) Reproduced with permission from [21], Copyright (2022) American Physical Society. (c,d) Reproduced with permission from [133], Springer Nature.

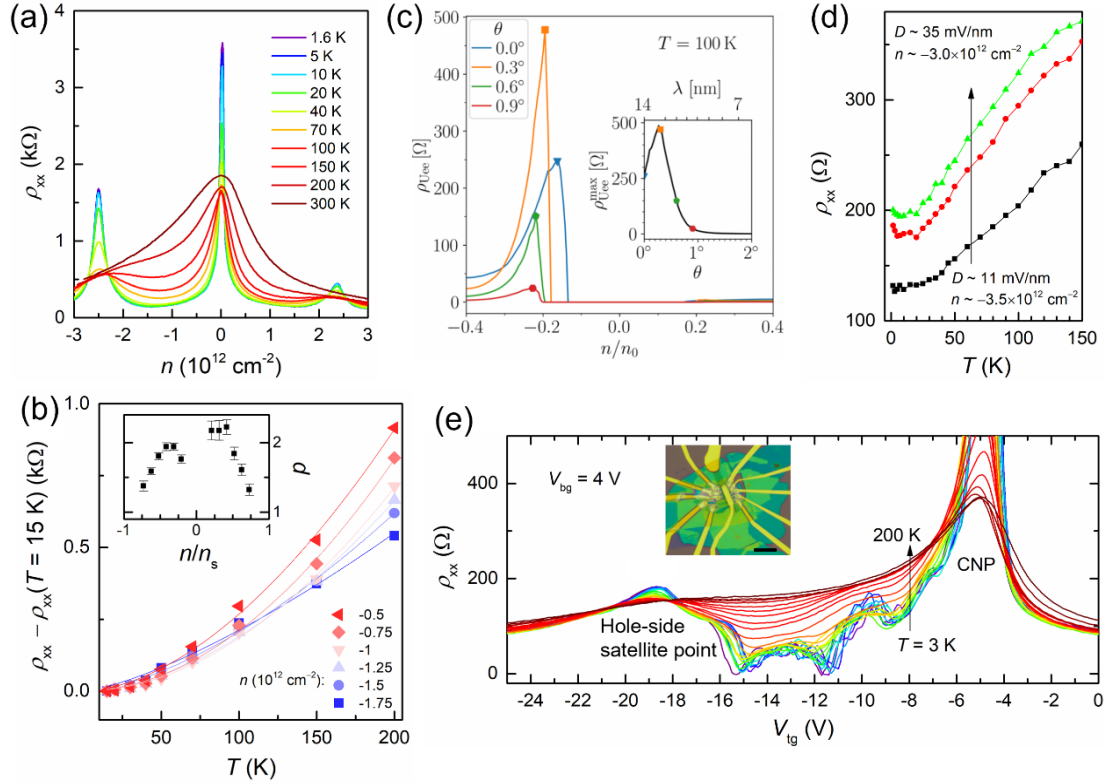

**Fig. 13. Resistance-temperature characteristics away from the CNP and satellite points.** (a) Longitudinal resistivity of device D1 as a function of carrier density for various temperatures. (b) Excess resistivity ($\rho_{xx}(T) - \rho_{xx}(T = 15\text{ K})$) of D1 as a function of temperature. The carrier density is set between the CNP and the hole-side satellite point. The symbols exhibit the experimental data, while the solid lines indicate the fitting results with the relation $\sim aT^p$ ($a$ is a parameter). Inset: Exponent ($p$) of the power-law dependence as a function of the carrier density normalized by the moiré full filling. (c) Contribution of the Umklapp effect to the resistivity as a function of carrier density for various stacking angles. Inset: Peak value of the resistivity induced by the Umklapp effect as a function of stacking angle. (d) Longitudinal resistivity of D1 as a function of temperature for various carrier densities and displacement fields. At low temperatures, the resistivity slightly increases under the specific conditions. (e) Longitudinal resistivity of another device as a function of top-gate voltage for various temperatures. A resistance bump and drop are observed between the CNP and the hole-side satellite point in the low-temperature regime. Inset: Optical image of the device. The scale bar corresponds to 10 μm. (c) Reproduced from [64], Copyright (2023) American Physical Society. (e) is modified and reproduced from [21].